\documentclass[twocolumn,preprintnumbers,amsmath,amssymb]{revtex4}

\usepackage{graphicx}
\usepackage{dcolumn}
\usepackage{bm}

\begin{document}

\title{Observation of breather-like states in a single Josephson cell}

\author{F. Pignatelli and A. V. Ustinov}
 \affiliation{Physikalisches Institut III, Universit\"at Erlangen-N\"urnberg, Erwin-Rommel-str. 1, D-91058 Erlangen, Germany.}
 
\date{\today}

\begin{abstract}
We present experimental observation of broken-symmetry states in a superconducting loop with three Josephson junctions. These states are generic for discrete breathers in Josephson ladders. The existence region of the breather-like states is found to be in good accordance with the theoretical expectations. We observed three different resonant states in the current-voltage characteristics of the broken-symmetry state, as predicted by theory. The experimental dependence of the resonances on the external magnetic field is studied in detail.
\end{abstract}

\maketitle

 \section{Introduction}
 
 Nonlinearity and discreteness in lattices lead to the existence of spatially localized stable solutions \cite{aubry,takeno,flach}. These solutions, called discrete breathers (DBs), are characterized by time periodicity and a strong spatial localization of energy, on the scale of the lattice constant. The localized mode can be vibrational or rotational, depending on the lattice and type of excitation. In the second case the DB is called \emph{rotobreather}. DB were found to exist in a large variety of nonlinear discrete lattices including weakly coupled optical wave guides \cite{eisenberg}, crystal lattice in solids \cite{swanson}, antiferromagnets \cite{schwarz} and Josephson junction arrays \cite{trias,peter,peter2,trias2,marcus}. The latter system is a convenient experimental tool to investigate the existence and dynamics of DBs. The discrete lattice here is constituted by an array of coupled underdamped small Josephson junctions (JJs).
 
 A JJ is composed by two superconducting layers separated by a thin tunnel barrier. Its dynamics can be described in the resistively-capacitively-shunted-junction model (RCSJ) \cite{bar}. The equation for the superconducting phase difference between the two superconductor layers constituting the junction, $\varphi$, is written in the normalized form as

\begin{eqnarray}
 \gamma=\ddot\varphi+\alpha\dot\varphi+\sin\varphi.
\end{eqnarray}
 Here the time unit is normalized by the inverse of the Josephson plasma frequency $\omega_{p}$, $\gamma$ is the bias current $I$ normalized by its critical value and $\alpha$ is the damping. This equation is formally equivalent to the equation of motion of a weakly damped pendulum in presence of an external torque $\gamma$. This mechanical analog allows for an easy description of the two possible states of the JJ. If we apply a torque $\gamma>1$ the pendulum will start to rotate with an average frequency $<\dot\varphi>$. After the pendulum is driven to the rotating state, it will stay in this state, even for $\gamma<1$, as long as the external torque is large enough to compensate the damping term. So that at some range of $\gamma$ both static and rotating states are possible, depending on the initial conditions. By the second Josephson equation $V=\Phi_{0}/2\pi<\dot{\varphi}>$, these two states are the equivalent, respectively, to the superconducting and resistive states of the JJ. In experiment by measuring the voltage on the JJ it is possible to distinguish between these two states.
 
 Josephson ladders allow for the existence of both oscillatory and rotational localized modes \cite{floria}. However only the latter ones are easy to detect experimentally. A rotobreather in the Josephson ladder is constituted by some of the JJs in the resistive (rotating) state, while the rest of JJs stays in the superconducting (oscillating) state.  The rotobreathers were experimentally detected by measuring the $I-V$ characteristics of the JJs in different points of the ladder \cite{trias,trias2} and also directly visualized by low temperature scanning laser microscopy \cite{peter,peter2}. 
 
 Josephson ladders can support linear cavity modes that can interact with the localized modes \cite{trias2,marcus,andrey,misha2}. The interaction of the linear modes of the ladder with the localized modes was found to lead to nonlinear features in the characteristic of DBs. Their complex dynamic in the region of the linear cavity resonances was investigated experimentally by the measurements of the $I-V$ characteristics of various DBs \cite{marcus}.

 Single Josephson cells have been proposed \cite{mazo,abdel} to provide an insight into complex discrete breather states of larger systems such as Josephson ladders and two dimensional Josephson junction arrays. Broken-symmetry states with characteristics similar to those of discrete breathers were proposed to exist \cite{mazo,abdel,misha}. We will call in the following these broken-symmetry states as \emph{breather-like} states. 
 
 Superconducting loops with three Josephson junction have also been recently proposed for the realization of a qubit \cite{orlando,yukon}. Such a loop has an important property required for the qubit, namely two closely positioned minima in the dependence of its energy on the external magnetic field \cite{yukon2}. Note, however, that in this work we are interested into phase-rotational states, which are clearly different from oscillatory states discussed for qubits.
 
\begin{figure}[h]
\includegraphics{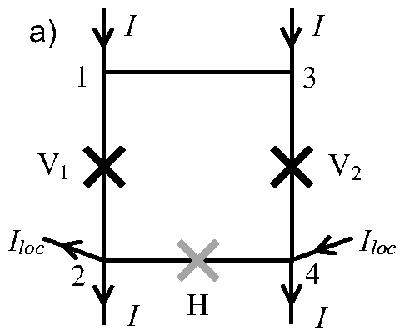}
  \includegraphics[width=3.3cm]{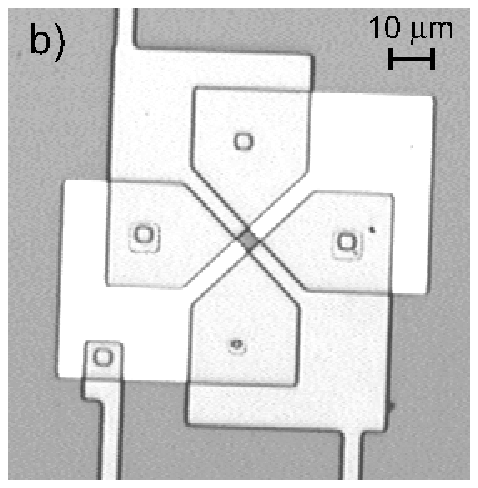}
\caption{\label{fig:cella} a) Sketch of the single Josephson cell, with three JJs (indicated by crosses), two vertical V$_{1}$, V$_{2}$ and one horizontal H. b) Optical image of our experimental cell with $4\times4$ $\mu$m$^{2}$ hole and three underdamped Nb/Al-AlO$_{x}$/Nb JJs.}
\end{figure}

 In this paper, we study a single Josephson cell  with three small underdamped Josephson junctions. Two junctions of same area placed along the direction parallel to the uniform bias current, V$_{1}$ V$_{2}$, are called vertical junctions. The third junction located in one of the two transverse branches of the cell, is called horizontal H, see Fig.~\ref{fig:cella}. 

\begin{figure}[h]
\includegraphics{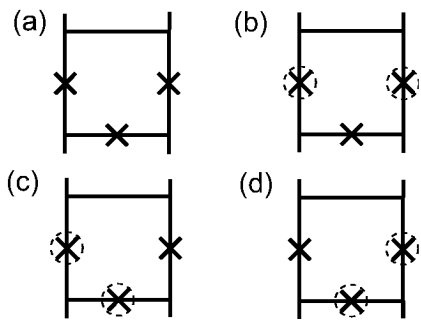}
\caption{\label{fig:states}Possible states of the system, (a) superconducting state, (b) homogeneous whirling state, (c) (d) the two symmetric breather-like states}
\end{figure}

 A parameter of cell \emph{anisotropy}, $\eta$, is defined as the ratio between the critical current of the horizontal junction and the one of the vertical junctions, $\eta=I_{CH}/I_{CV}$. The parameter of anisotropy describes the coupling between the two vertical junctions, so that in the limit $\eta\to 0$ the two vertical junctions are entirely decoupled, while the larger is $\eta$ the stronger is the coupling. The cell is called isotropic if all three junctions have same parameters, anisotropic otherwise. The system is uniformly biased by two equal currents $I$, applied through the vertical branches, 1-2 and 3-4, Fig.~\ref{fig:cella}. As long as the uniform bias current $I$ is smaller than the critical current $I_{CV}$, all the junctions stay in the superconducting state, Fig. \ref{fig:states}a. As the current $I$ exceeds the critical value $I_{CV}$, the system switches to the homogeneous whirling state, constituted by both vertical junctions in the resistive state and the horizontal one in the superconducting state, Fig. \ref{fig:states}b. Both these cases are symmetric states and there is no current on the horizontal junction, so that the two vertical junctions behave identically. We are interested in the last two cases, Fig.~\ref{fig:states}c and d, which are the inhomogeneous breather-like states. These two states have the same behavior and are characterized by the horizontal junction and one of the vertical junctions in the resistive state, while the other vertical junction remains in the superconducting state. In these cases, obviously, some part of the bias current flows through the transverse branch. 

 The procedure to generate breather-like states in experiments is similar to that used for Josephson ladders \cite{trias,peter}. We bias the system by a local current, $I_{loc}$, through the transverse branch 4-2, Fig.~\ref{fig:cella}, and rise this current until the system switches to a breather-like state. Afterwards, we increase the bias current $I$ and simultaneously decrease $I_{loc}$ to zero, keeping the system in the generated state. Alternative procedure is first apply a homogeneous bias current $I<I_{CV}$ at a value where we expect to find the breather-like state stable, then rise the local current $I_{loc}$ to generate it, and finally reduce $I_{loc}$ back to zero. By using the two different directions of the initial local current it is possible to generate both breather-like states shown in Fig.~\ref{fig:states}. 
 
 Different states of the system can be experimentally detected by measuring the dc voltage on  Josephson junctions. As for Josephson ladders, the broken-symmetry states are the consequence of the characteristic property of small underdamped Josephson junctions, that is the coexistence, for some ranges of the bias current, of two different stable states, superconducting and resistive.

 \section{Dynamics of the system} 
 The dynamics of the system are described by the phase differences across the three Josephson junctions, $\varphi_{V_{1}}$, $\varphi_{V_{2}}$, $\varphi_{H}$. Using the current conservation law in each node of the cell and the flux quantization for the cell, one can get the equations of motion for the system \cite{abdel,misha}:

\begin{eqnarray}
\mathcal{N}(\varphi_{V_{1}}) &= &\gamma-\frac{1}
{\beta_{L}}(\varphi_{V_{1}}-\varphi_{V_{2}}+\varphi_{H}+2\pi f)
\nonumber
\\
\mathcal{N}(\varphi_{V_{2}}) &= &\gamma+\frac{1}
{\beta_{L}}(\varphi_{V_{1}}-\varphi_{V_{2}}+\varphi_{H}+2\pi f)
\label{motion}
\\
\mathcal{N}(\varphi_{H}) &= &-\frac{1}
{\beta_{L}\eta}(\varphi_{V_{1}}-\varphi_{V_{2}}+\varphi_{H}+2\pi f),
\nonumber
\end{eqnarray}
\\
 where the time is normalized by the inverse of the Josephson plasma frequency $\omega_{p}$. The influence of the external magnetic field is introduced by the term $2\pi f/\beta_{L}$. The parameters of the system are the uniform \emph{bias current} normalized by the critical current of the vertical junctions, $\gamma=I/I_{CV}$, the normalized \emph{self-inductance} of the cell, $\beta_L=2\pi L I_{CV}/\Phi_{0}$ (where $L$ is its real inductance), the \emph{damping}, $\alpha=\sqrt{\Phi_{0}/(2\pi I_{C}R^{2}C)}$, and the frustration parameter $f=\Phi_{ext}/\Phi_{0}$, that can be tuned by the external magnetic field $H$. The operator $\mathcal{N}$ represents the current through a single junction in the RCSJ model and is defined as: $\mathcal{N}(\varphi)=\ddot\varphi+\alpha\dot\varphi+\sin\varphi$. 
 
 Using a simplified dc model \cite{abdel}, where the junctions in the resistive state are considered as resistances and the ones in the superconducting state just as shorts, one can estimate the dimensionless breather frequency $\Omega=<\dot{\varphi}>$, that is in fact the voltage across a junction in the resistive state normalized by $\Phi_{0}\omega_{p}/2\pi$:

\begin{eqnarray}
\Omega = \frac{\gamma}{\alpha(1+\eta)}.
\label{DB}
\end{eqnarray}
 From that one can also evaluate the region of existence of the breather-like state \cite{abdel}. The minimum current at which a broken-symmetry state may exist is

\begin{eqnarray}
\gamma_{r}=\frac{4\alpha}{\pi}(1+\eta).
\label{min}
\end{eqnarray}
 At $\gamma<\gamma_{r}$ the system is retrapped to the superconducting state. The maximum current for a breather-like state is
 
\begin{eqnarray}
\gamma_{c}=\frac{1+\eta}{1+2\eta}.
\label{max}
\end{eqnarray}
 Above this current the system switches to the homogeneous rotating state.

 The simplified dc model neglects the influence of the small oscillations of the Josephson phase that may cause an instability and nonlinearity in the characteristic of the breather-like state. Neglecting the damping and linearizing the equation of motion, Eqs.~(\ref{motion}), around the breather-like state, Eq.~(\ref{DB}), it is possible to calculate the two characteristic frequencies of these oscillations \cite{abdel,misha}:

\begin{eqnarray}
\omega_{\pm}=\sqrt{F\pm\sqrt{F^{2}-G}},
\label{res}
\end{eqnarray}
\begin{eqnarray}
\textrm{where:}&F&=\frac{1}{2}\cos(c_{1})+
\frac{1+2\eta}{2\eta\beta_L}
\nonumber
\\
&G&=\frac{1+\eta}{\eta\beta_L}\cos(c_{1}).
\nonumber
\end{eqnarray}
\\
 Here $c_{1}$ is the phase shift of the vertical junction in the superconducting state evaluated by the dc model, $c_{1}=\arcsin[\gamma(1+2\eta)/(1+\eta)]$. The presence of the electromagnetic oscillations leads then to \emph{primary} resonances in the breather-like state, when the frequency of the breather $\Omega$ locks to the frequency of the electromagnetic oscillations $\omega_{+}$ or $\omega_{-}$, and to \emph{parametric} and \emph{combination} resonances, respectively, at $\Omega=n\omega_{\pm}$, where $n\ge2$ is an integer, and $\Omega=n(\omega_{+}\pm\omega_{-})$, where $n$ is an integer \cite{abdel,misha}.
 
 Using the rotational approximation \cite{kulik}, two different dependencies of the resonant step amplitude $\Delta\gamma$ on the frustration have been obtained for \emph{primary} and \emph{parametric} resonances \cite{misha}:

\begin{eqnarray}
\Delta\gamma_{\omega_{\pm}} &\propto& \bigg\vert\sin\bigg(\frac{c_{1}}{2} 
+\frac{\beta_{L}\gamma\eta}{2(1+\eta)}+\pi f\bigg)\bigg\vert
\label{prim}
\\
\Delta\gamma_{n\omega_{\pm}}& \propto& \sqrt{(1-\eta)^{2}+4\eta\cos^{2}
\bigg(\frac{c_{1}}{2}+\frac{\beta_{L}\gamma\eta}{2(1+\eta)}+\pi f\bigg)}.
\nonumber
\\ \label{para}
\end{eqnarray}
 Expressions (\ref{prim}) and (\ref{para}) have the same periodicity in $f$ as the critical current in the whirling state, but, depending on the parameters of the system, they may be asymmetric with respect to $f=0$. Moreover, the external magnetic field can considerably change the amplitude of the resonance and, in the case of primary resonance, even make it to vanish for some values of the frustration.
 
 \section{Experimental results}
 Our experimental system is a single cell with the hole area $4\times4$ $\mu$m$^{2}$, containing three small underdamped Nb/Al-AlO$_{x}$/Nb Josephson junctions \cite{hypres}, see Fig. \ref{fig:cella}. The critical current density, measured at 4.2 K, is $J_{C}=133$ A/cm$^{2}$. The area of each vertical junction is about $7\times7$ $\mu$m$^{2}$, while the horizontal one is $3\times4$ $\mu$m$^{2}$. The anisotropy of the system, evaluated from the design, is then $\eta=0.24$. The parameter of self-inductance of the cell at $4.2$ K, evaluated by measurement of a SQUID of same geometry, is $\beta_{L}=1.47$. The two equal bias currents are introduced in parallel using two large external resistances of $1.5$ k$\Omega$, so that the bias current remains uniform also in the broken-symmetry states. The damping of the system at 4.2 K is evaluated from the retrapping current of the homogeneous whirling state, $\alpha\simeq(\pi/4)(I_{r}/2I_{CV})$, that yields $\alpha\simeq 0.03$. The measurements are performed in a $^{4}$He cryostat \cite{cryo} in the temperature range between 4 K and 8 K, so that the parameters of the system $J_{C}$, $\alpha$, $\beta_{L}$ and $\omega_{p}$ can be varied. For each temperature the damping $\alpha$ and the critical current density $J_{C}$ are evaluated from the characteristic of the homogeneous whirling state, while the parameter of self inductance $\beta_{L}$ is evaluated from the critical current density. Finally, the value of the plasma frequency is obtained at every temperature from fitting the voltage of the parametric resonant step to the expected value $2\omega_{+}$.
 
 For convenience of comparing with theory, the experimental data are shown in the previously introduced normalized units of the bias current $\gamma=I/I_{CV}$,  breather frequency $\Omega=V/(\omega_{p}\Phi_{0}/2\pi)$ and frustration $f=\Phi_{ext}/\Phi_{0}$.
 
 In the range of temperatures from 4.2 K to 7.75 K ($0.02\le\alpha\le0.35$) it was possible to excite both breather-like states and they were found stable in a finite range of the bias current. No breather-like state was found stable for temperatures above 7.75 K ($\alpha>0.35$).    
 
\begin{figure}[!htb]
  \includegraphics[width=6cm,angle=-90]{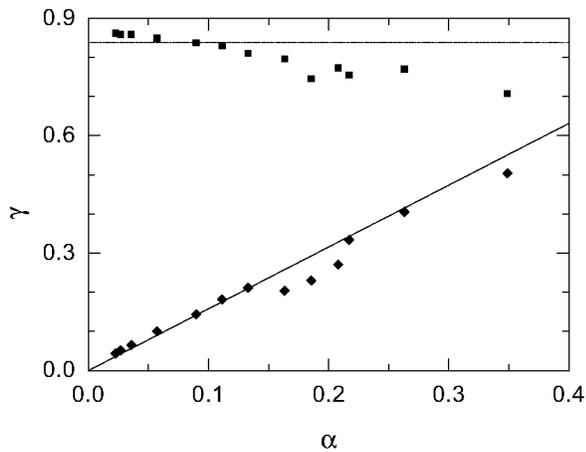}
  \caption{\label{fig:exist}Maximum (solid squares) and minimum (solid diamonds) currents for the breather-like state. The solid lines show the theoretical expectation of the dc model, Eqs. (\ref{min}) and (\ref{max}).}
\end{figure}
 
 The minimum bias current below which the breather is retrapped increases linearly with the damping, see Fig.~\ref{fig:exist}, in good agreement with the theoretical expectation of Eq~(\ref{min}). The maximum current that allows for the existence of the breather-like state slightly decreases with increasing the damping, see Fig. \ref{fig:exist}, whereas from Eq. (\ref{max}) it is expected to be constant ($\gamma_{c}=0.84$). 
 
 \begin{figure}[!htb]
   \includegraphics[width=6cm,angle=-90]{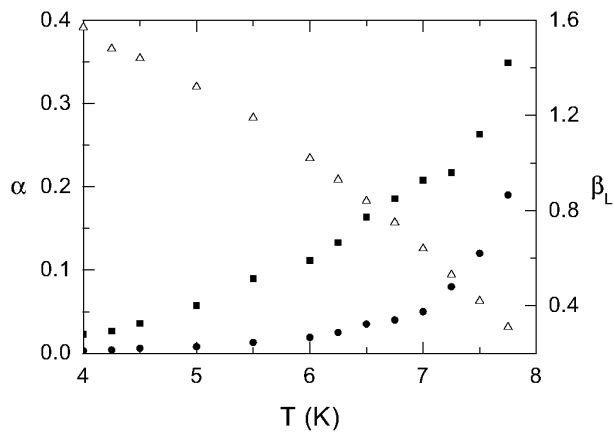}
   \caption{\label{fig:parameters}Dependence of the damping, evaluated by the retrapping current (solid squares) and by the sub-gap resistance (solid circles), and of the self-inductance (open triangles) on the temperature of the system.}
\end{figure}
 
 In fact, the temperature variation leads to an alteration of several parameters, see Fig.~\ref{fig:parameters}. In particular the decrease of the self inductance $\beta_{L}$ at high temperatures leads to stronger interactions of the breather-like state with the small oscillations. This effect may strongly affect (in the range $0.1<\beta_{L}<1$)  the region of existence of the breather-like, especially its upper border \cite{abdel}. In fact, the experimental data show deviation from the expected value for $\alpha\ge 1.15$, which is in the range of the self inductance between 0.9 and 0.3.
 No resonances were observed in the $I-V$ characteristics at temperatures below 6 K. The small damping of the junctions at this temperatures seems to be unfavorable for reaching the small frequencies of the electromagnetic oscillations where the resonances are expected ($\Omega_{min}\simeq5.5$). Increasing the temperature of the system had two helpful effects for seeing the resonances, namely the increase of damping and decrease of self inductance, see Fig.~\ref{fig:parameters}. As the consequence of the larger damping, the breather-like states were stable down to smaller frequencies, whereas a smaller inductance moved the frequencies of the resonances up.
 
\begin{figure}[!htb]
 \includegraphics[width=6.3cm,angle=-90]{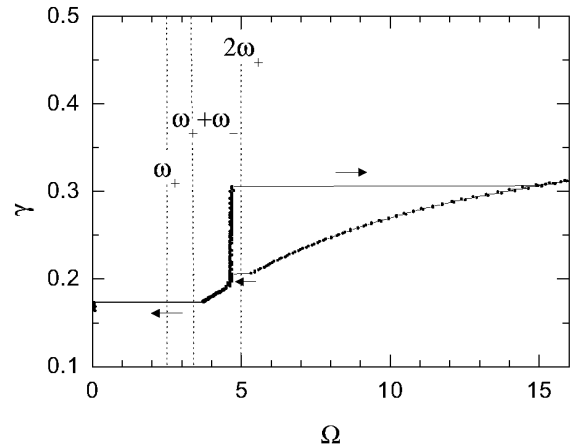}
 \caption{\label{fig:6k}Measured current-voltage characteristic of the breather-like state at damping $\alpha\simeq 0.11$ and frustration $f=-0.1$, where the magnitude of the resonant step is maximum. The dotted lines show the expected frequencies of the electromagnetic oscillations, Eq.~(\ref{res}).}
\end{figure}

 At the temperature of $T=6$ K it was  possible to observe the first resonant step. The parameters of the system at this temperature are $\alpha\simeq 0.11$, $J_{c}=92$ A/cm$^{2}$,  $\beta_{L}=1.02$ and $\omega_{p}/2\pi=37$ GHz. This resonance is a step in the characteristic, shown in Fig.~\ref{fig:6k}. Comparing this resonance with that  at higher temperatures and with the theoretical expectation we deduce it to be the parametric resonance at $2\omega_{+}$.

\begin{figure}[!htb]
 \includegraphics[width=6.3cm,angle=-90]{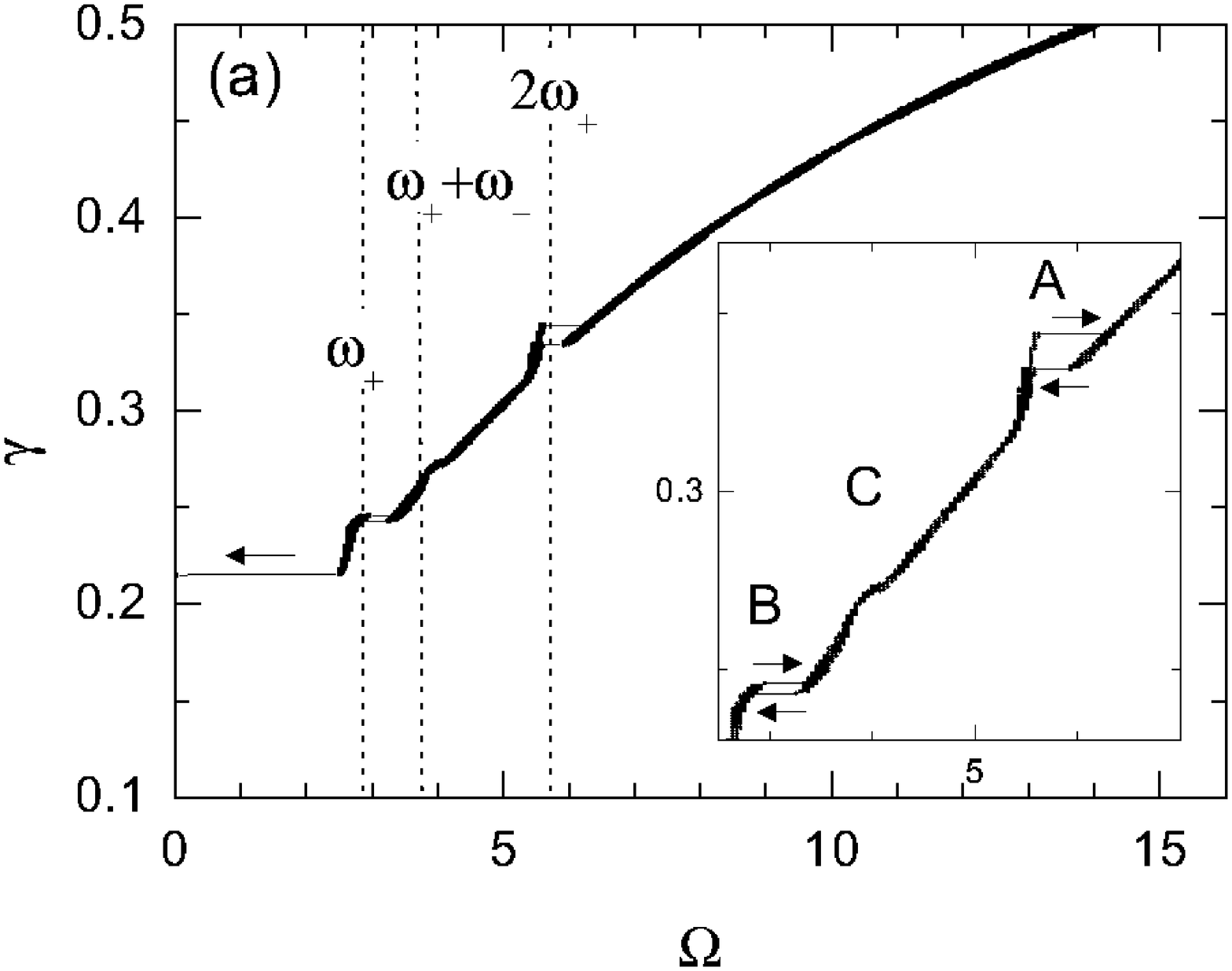}
 \includegraphics[width=6.3cm,angle=-90]{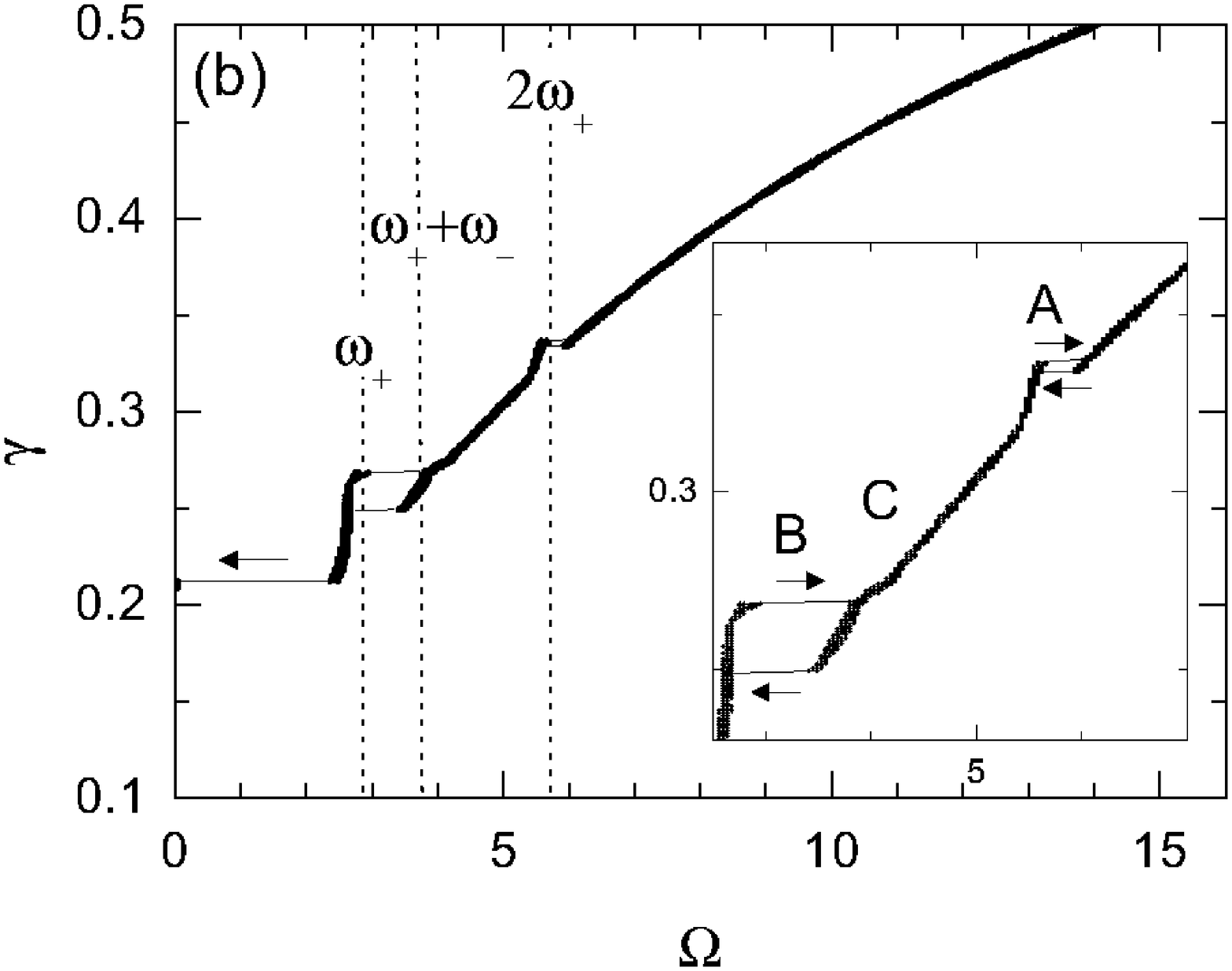}
 \caption{\label{fig:reso1}Measured current-voltage characteristic of the breather-like state at the damping $\alpha\simeq 0.18$ and for:(a) $f=-0.1$, (b) $f=0$. The dotted lines show the expected frequencies of the electromagnetic oscillations, Eq.~(\ref{res}). The insets show a zoom into the region of interest.}
\end{figure}

 At the temperature of $T=6.65$ K we detect two additional resonances in the characteristic of the breather-like state, see Fig.~\ref{fig:reso1}. At this temperature the parameters of the system are $\alpha\simeq 0.18$, $J_{c}=70$ A/cm$^{2}$,  $\beta_{L}=0.77$ and $\omega_{p}/2\pi=30$ GHz. The two larger resonances, A and B, are different by the integer factor of two, as expected for $2\omega_{+}$ and $\omega_{+}$.

\begin{figure}[!htb] 
 \includegraphics[width=6.3cm,angle=-90]{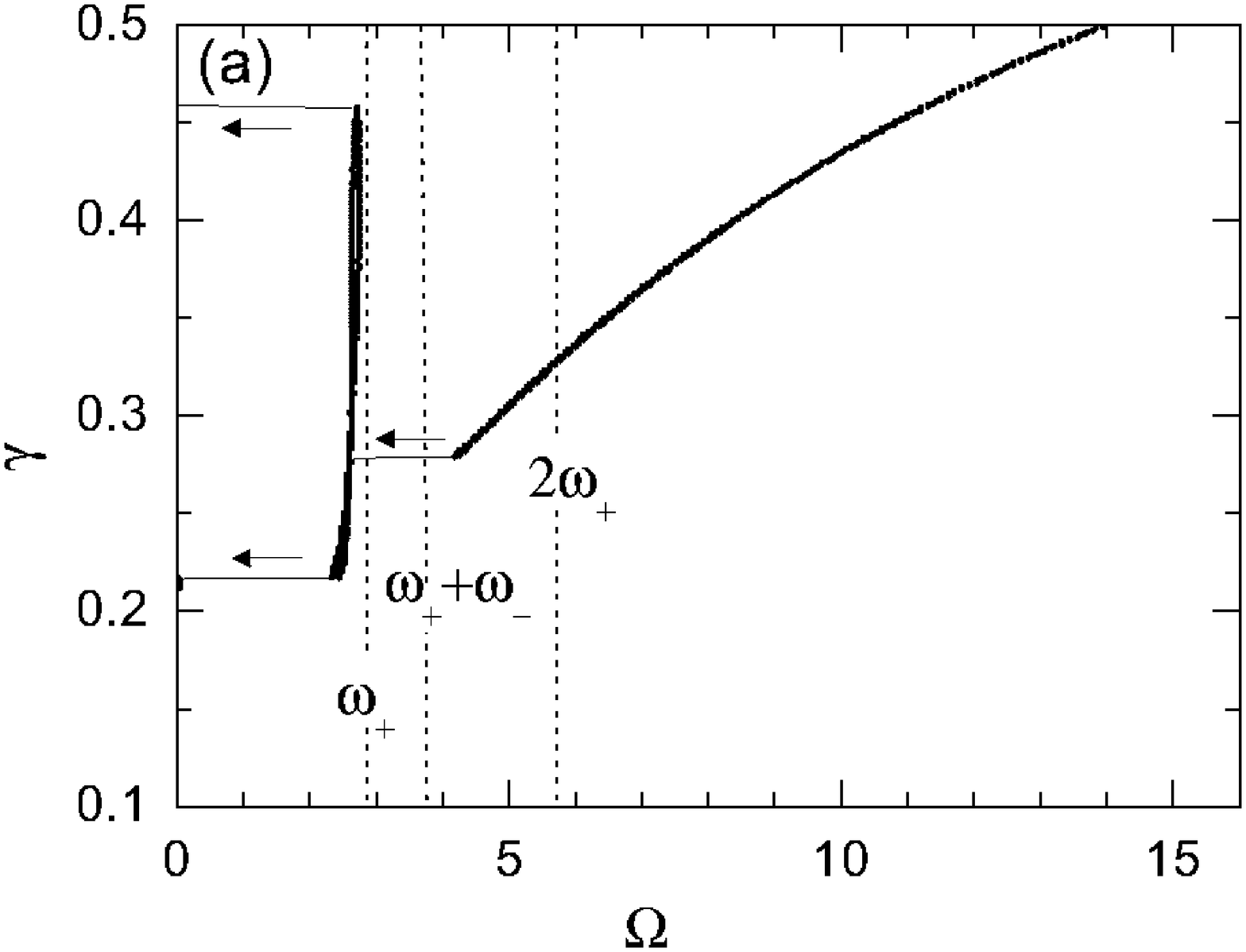}
 \includegraphics[width=6.3cm,angle=-90]{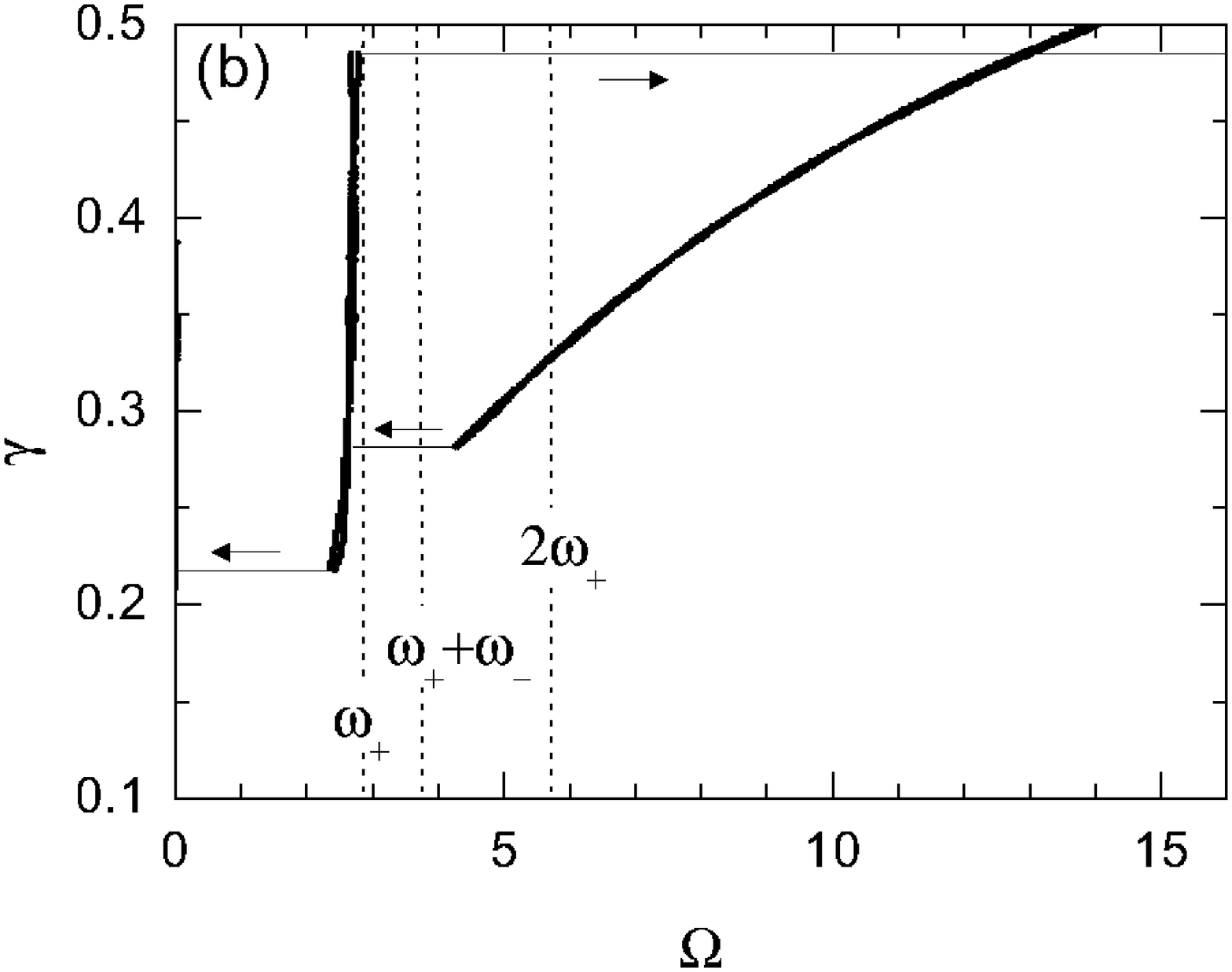}
\caption{\label{fig:reso2}Measured current-voltage characteristic of the breather-like state in presence of instabilities at damping $\alpha\simeq 0.18$ (a) $f=-0.6$, (b) $f=0.4$. The dotted lines show the frequencies of the electromagnetic oscillations evaluated from Eq.~(\ref{res}).}
\end{figure}

 The dependencies of the current amplitude on the external magnetic field for steps A and B are strongly different. The resonance A has its maximum amplitude close to zero frustration, see Fig.~\ref{fig:reso1}a, and disappears in a wide range around $f=0.5$, see Fig.~\ref{fig:reso2}. An other resonance B is always present but has nearly opposite behavior. At the top of the resonance B, around frustration $f=0.4$ or $f=-0.6$, the system shows an instability by switching either to the homogeneous whirling state or to the superconducting state, as shows Fig.~\ref{fig:reso2}. In this range of frustration the measurements of the resonant step B were performed with less accuracy and faster because of its poor stability. We argue the larger step B is due to the primary resonance at $\Omega=\omega_{+}$, whereas the step A is due to the parametric resonance at $\Omega=2\omega_{+}$. The third smaller nonlinearity, C, is at a frequency between the other two, close to the theoretically expected value for the composed resonance at $\Omega=\omega_{+}+\omega_{-}$, see Fig~\ref{fig:reso1}. This resonance was only present in a small range of frustration around zero, where from the theory, the composed resonance is expected to be maximum. Because of its reduced amplitude this resonant step has no hysteresis and its dependence on the magnetic field was not measured. 
 
\begin{figure}[!hbt]
 \includegraphics[width=5.7cm,angle=-90]{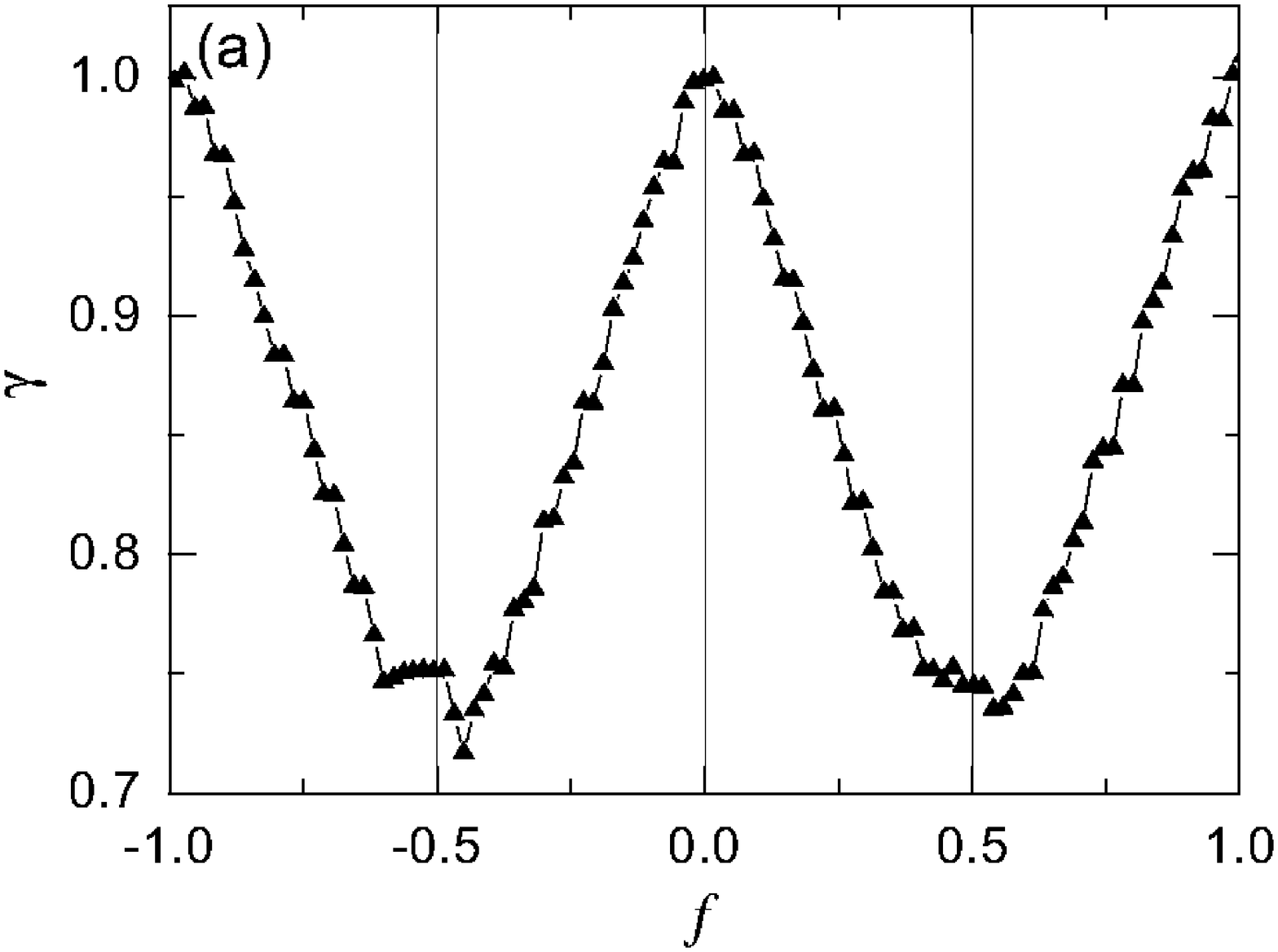}
 \includegraphics[width=5.7cm,angle=-90]{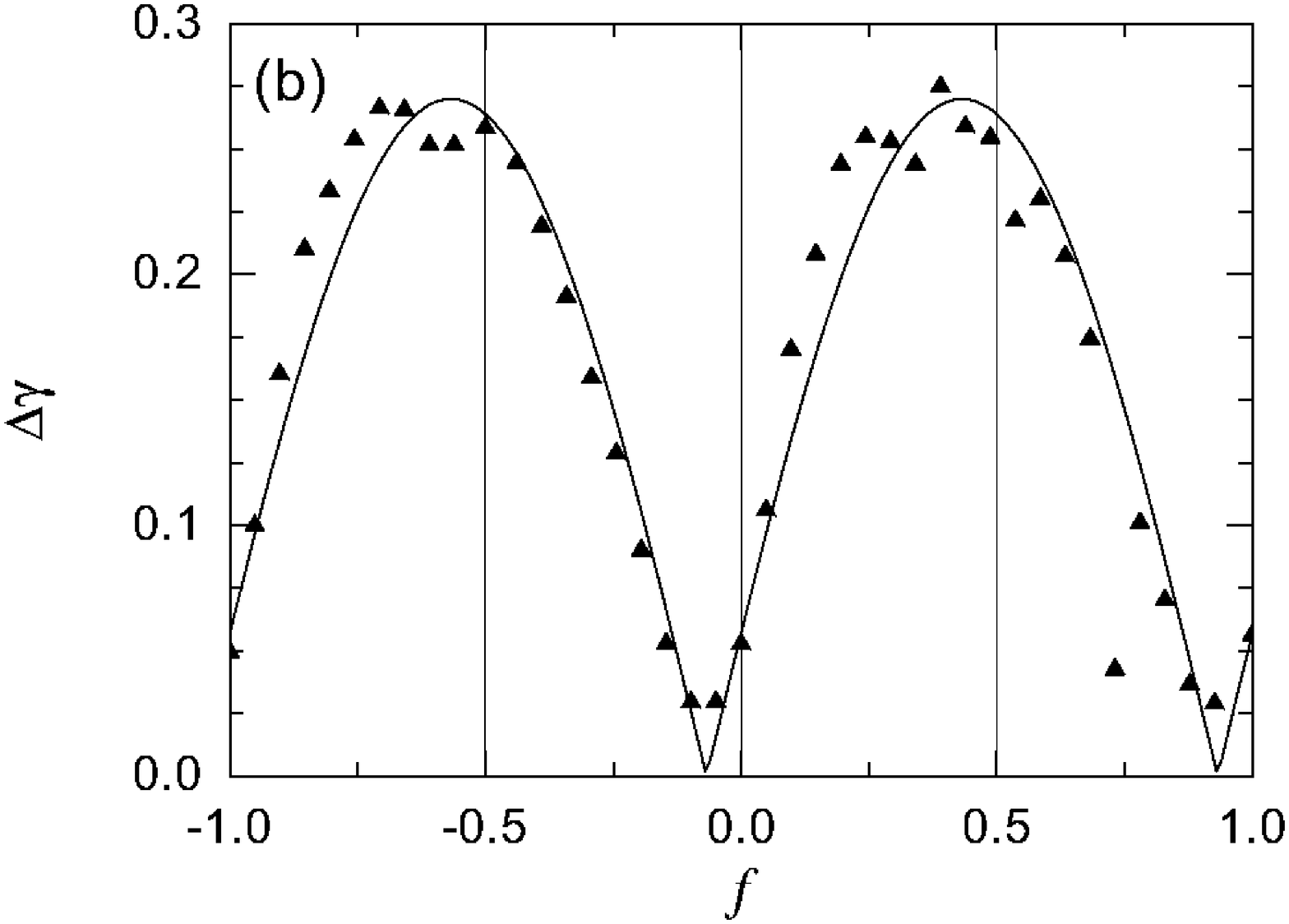}
 \includegraphics[width=5.7cm,angle=-90]{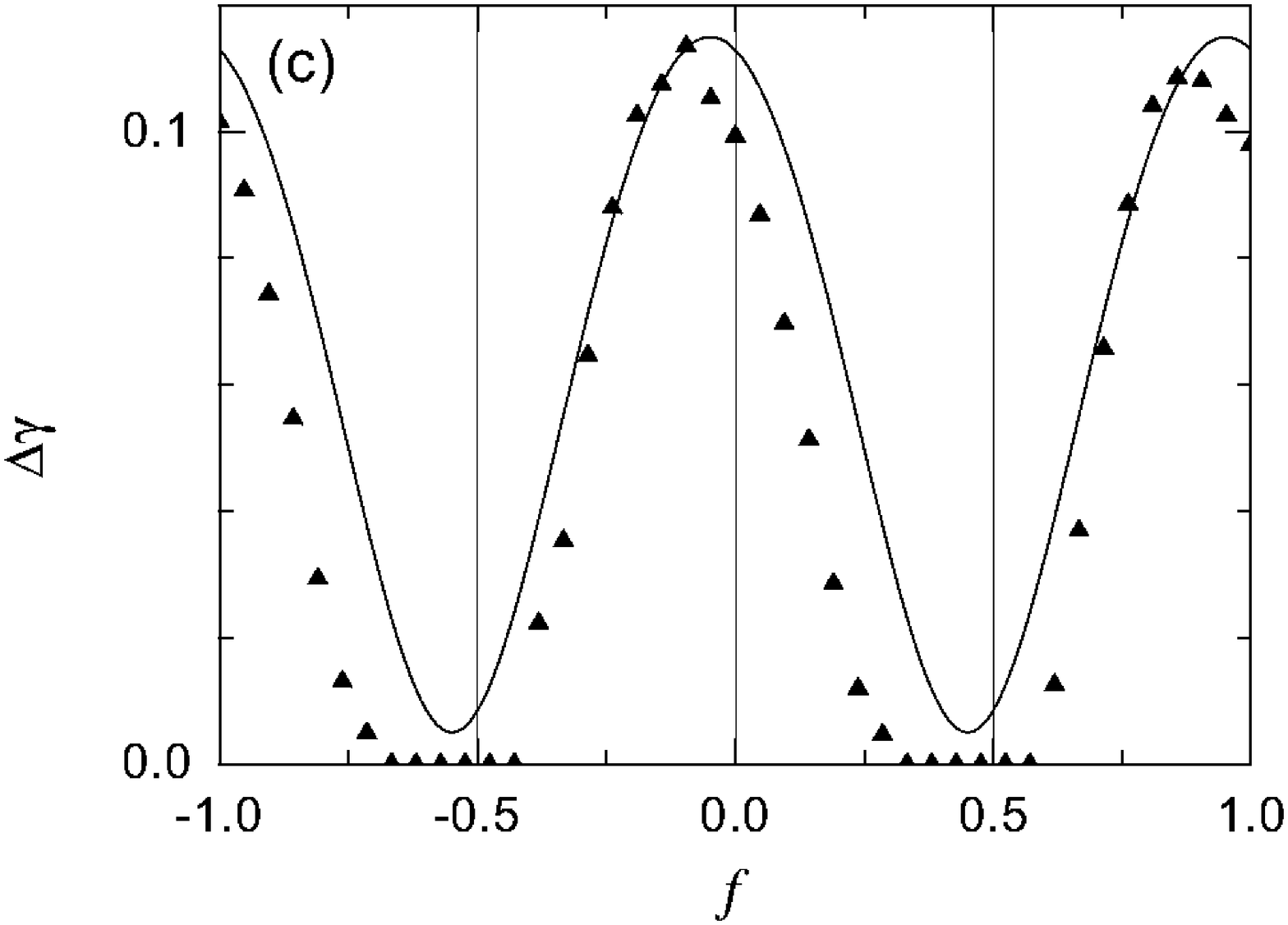}
 \caption{\label{fig:frustr}(a) Dependence of the critical current of the homogeneous state on the frustration. (b), (c) Dependence of the magnitude of the resonant step on the frustration for primary resonance A (measured at $T=6.65$ K) and parametric resonance B (measured at $T=6$ K) respectively. The solid lines are the theoretical curves given by Eq.~(\ref{prim}) and Eq. (\ref{para}), respectively.}
\end{figure}

 In Fig.~\ref{fig:frustr}b, c we present measurements of the current amplitude of the resonant steps A and B on the external magnetic field. The parametric step at $T=6$ K, see Fig.~\ref{fig:6k}, is higher than that at $T=6.65$ K, Fig.~\ref{fig:reso1}. Moreover, because of the larger damping in the latter case the step vanishes in a larger range of the frustration ($-0.95<f<-0.25$ and $0.05<f<0.75$). Therefore, in order to observe a more clear modulation of this resonance, we report in Fig.~\ref{fig:frustr}c its dependence on the external magnetic field at $T=6$ K.

 As expected from the theory, Eq.~(\ref{prim}) and Eq. (\ref{para}), the magnetic field dependence of the resonance amplitudes show an asymmetry with respect to $f=0$. Note, that such asymmetry is not observed for $I_{C}(f)$, see Fig.~\ref{fig:frustr}(a). The asymmetry depends on the parameter of the system and particularly on the bias current where the resonance occurs. This behavior can be qualitatively attributed to the current flowing in the transverse branch of the cell in the broken-symmetry state. The experimental data are in rather good agreement with the theoretical expectations. We fitted the maximum and minimum of the resonant steps in order to find the proportionality coefficient in Eqs.~(\ref{prim}) and (\ref{para}). The small variations of the experimental data around the maximum of the primary resonant step could be due to the appearance of instabilities, as discussed in relation to Fig.~\ref{fig:reso2}. This behavior was also found in simulations \cite{misha}. In the case of parametric resonance the presence of a range of frustration, that increases with the damping $\alpha$, where resonant step disappears, is also expected \cite{misha}. Note, that the region where the experimental data show the largest deviation from the expected behavior is nearly the same for the two cases ($0.3<f<0.6$).
 
 \section{Conclusions}

 We observed breather-like states of broken symmetry in an anisotropic single-cell system. For temperatures lower than 6 K the current-voltage characteristics of the breather-like states were stable and did not show any resonances. Increasing the dissipation by rising the temperature of the system we decreased the frequencies of rotation of the breather-like state in such a way that it matched the frequencies of the resonances in the cell. In this region, the primary, composed and parametric resonances were observed. As expected, in all three cases, the interaction of the electromagnetic oscillations with the breather-like state leads to resonant steps in the current-voltage characteristic. The large steps due to the primary and parametric resonances show hysteresis, whereas the composed one induces just a small nonlinearity in the characteristic. The measured dependence on the external magnetic field of the magnitude of the primary and parametric resonant steps is in good agreement with theory. In the case of primary resonance, the external frustration can strongly increase the amplitude of the resonant step and lead to instability of the breather-like state. The composed resonance, for this values of the damping, was present just in a small range around zero frustration and did not lead to instabilities. The region of existence of the breather-like state was also measured and found in good accordance with the expectations. At smaller values of the self inductance parameter, the stronger interaction of the breather with the electromagnetic oscillations leads to a contraction of the region of existence.

\section{acknowledgments}
We would like to thank A. Benabdallah, P. Binder, M. V. Fistul, S. Flach, A. E. Mirishnichenko and M. Schuster for the fruitful discussions and for sending us their latest works before publication. 
This work was supported by the Deutsche Forschungsgemeinschaft (DFG) and by the European Union under RTN project LOCNET HPRN-CT-1999-00163.

\end{document}